# Determination of InN-GaN heterostructure band offsets from internal photoemission measurements

Zahid Hasan Mahmood

*Department of Applied Physics, Electronics and Communication Engineering, University of Dhaka, Dhaka 1000, Bangladesh*

A.P. Shah, Abdul Kadir, M. R. Gokhale, Sandip Ghosh, Arnab Bhattacharya and B. M. Arora

*Department of Condensed Matter Physics and Materials Science, Tata Institute of Fundamental Research, Homi Bhabha Road, Mumbai 400 005, India*

**Abstract**

Band discontinuities at the InN-GaN herterointerface are experimentally determined from internal photoemission spectroscopy measurements on $n^+$ InN on GaN epilayers. The photocurrent shows two threshold energies, one at 1.624 eV and the other at 2.527 eV. From these we obtain the band offsets $\Delta E_v = 0.85$ eV and $\Delta E_c = 1.82$ eV.

**PACS No**. 73.21.–b, 73.40.Kp, 78.66.Fd, 79.60.Jv, 79.60.–i

**Key words**: InN-GaN heterostructures, Band offset, Internal photoemission

There is remarkable interest in InN because of its excellent theoretically predicted electrical properties and potential for application in high speed devices[1]. High-quality low band gap (0.7eV) InN layers could extend the wavelength range of optoelectronic devices covered by the group-III nitrides into the near infrared. With continual improvements in the growth quality of InN epitaxial layers, prospects of realising devices based on heterostructures have advanced considerably. This in turn has intensified efforts for accurately determining the basic properties of InN and its heterostructures. For example, intense experimental effort has led to near consensus on the value of the fundamental gap of good quality InN to be ~ 0.7 eV.[2,3] Effective mass of electrons has been determined.[4] Among the heterostructure properties, the most significant are discontinuities of band offsets at the junctions. The earliest estimate of the valence band offset at an InN-GaN heterojunction ($\Delta E_v$ = 0.48 eV) was obtained theoretically by Wei and Zunger using density functional theory.[5] Bhouri et al.[6] have theoretically examined the role of strain and of heterostructure ordering InN on GaN vs. GaN on InN on the band offsets in the zinc-blende phase and obtain $\Delta E_v$ values of 0.57eV and 0.77eV respectively. In contrast, the $\Delta E_v$ value obtained by Wei and Zunger for zinc-blende GaN/InN is 0.26eV. This clearly points to the need for more refined theories for zinc-blende and wurtzite phases. Early experimental measurements using X-ray photoelectron spectroscopy (XPS) by Martin et al. reported a value of $\Delta E_v$ ~ 1.05 eV.[7] A recent XPS measurement by Shih et al.[8] gives $\Delta E_v$ ~ 0.5 ± 0.1 eV close to the theoretical value of Wei and Zunger. Shih et al. point out that the larger $\Delta E_v$ value obtained in Ref. [7] could be related to inferior material quality. They suggest contamination with oxygen, indium clusters, or the presence of indium oxy-nitride to be responsible for the discrepancy. Considering these uncertainties and the precision of XPS measurements, it is desirable to determine band offsets by an alternative technique. We chose internal photoemission since this technique can provide accurate values of barrier heights.[9,10] From such measurements on high-quality epitaxial n$^+$ InN-GaN heterostructure epilayers, we determine $\Delta E_v$ close to 0.85 eV and $\Delta E_c$ is estimated to be about 1.82 eV. Details of the sample structure, measurements and results are reported in this letter.

The synthesis of InN via metalorganic vapor phase epitaxy (MOVPE) is the most difficult among the III–nitrides because of its low dissociation temperature and high equilibrium vapor pressure of N$_2$ over InN. The wurtzite InN epilayers studied in this work are grown in a close-coupled showerhead reactor system under optimized conditions arrived at by varying different MOVPE growth parameters, resulting in material of excellent crystalline quality. The details of the MOVPE growth of the InN epilayers have been published earlier.[11] Before the growth of the InN layer, a 1μm-thick undoped GaN buffer layer was grown on the *c*-plane sapphire substrate at 1040ºC using a standard two-step process. The temperature was then ramped down to 530ºC to grow a 0.2 μm-thick InN layer at reactor pressure of 300 Torr, and a V/III ratio of 18700. From Hall measurements in the Van-der-Pauw geometry the electron concentration in the InN was found to be ~ $10^{19}$ cm$^{-3}$. Although the GaN buffer is undoped, it has background electron concentration in the range few times $10^{15}$ cm$^{-3}$.

For the internal photoemission measurements 600 μm diameter mesas through the InN/GaN heterostructure were defined by photolithography and etched using reactive ion etching in a BCl$_3$ plasma. The mesa depth is controlled such that while the InN islands are completely isolated, the underlying GaN is connected over the whole sample. Following this, 400 μm diameter contacts on top of the InN mesa and large-area contact pads on the GaN encircling the InN mesa are photolithographically defined. Electron-beam evaporated Ni-Au is then used to form both contacts in a single-step metallization. The structures fabricated in this manner have rectifying I-V characteristics of nature similar to reported by Chen at al.[12] C-V characteristics give the doping concentration of GaN in the range 4x$10^{15}$ cm$^{-3}$ and a built-in voltage of about 1.1–1.2 V. The $1/C^2$–V characteristics are not quite linear and the estimated built-in voltage is an approximate value. The internal photoemission measurement set up has a quartz-tungsten-halogen lamp source chopped at 23 Hz, a 1/8 m Oriel monochromator with a grating blazed at 0.5 μm and a 430 nm long pass filter for order sorting. Most of the measurements have been made with a spectral band pass of 6.4 nm. Light is focused onto the mesa structure through the sapphire. The current is collected from contacts on the top of the InN mesa and on the GaN, and measured using a lock-in amplifier. The source light spectrum is calibrated by using a standard silicon diode and the measured response is normalized with respect to the source spectrum. All measurements have been done at room temperature.

Fig. 1(a) shows normalized photoresponse of the heterostructure over photon energies of 1.5 eV to 3.5 eV. As seen the response has three regions, marked I,II and III. In region I, the response is linear with respect to the photon energy, while in region II, it is non linear. We attribute the response I to photoexcitation of electrons from conduction band of InN and subsequent emission into GaN (process $E_1$, in the schematic band diagram in Fig. 2). The region II represents the photoelectron current due to excitations from the valence band of InN in addition to the excitations from the InN conduction band (process $E_2$, in the schematic band diagram in Fig. 2). Region III is the characteristic photocurrent spectrum from near the band edge of GaN, which was measured after removing the 430 nm long pass filter from the light path.

It is well known[9] that a linear response is characteristic of excitations and emission from a narrow band as in the case of partially-filled conduction band of InN. On the other hand, response from the valence band excitations is expected to follow a quadratic behaviour. Thus the total response can be represented as

$$Y = C_1(h\nu - E_1) + C_2(h\nu - E_2)^2$$

Where $E_1$ represents the emission threshold for transitions from the Fermi energy to the GaN conduction band edge, and $E_2$ is the emission threshold from the InN valence band edge to the GaN conduction band edge. In order to determine $E_1$, we have replotted in Fig. 1(b) the linear region I. The straight line drawn through the experimental points gives $E_1 = 1.624$ eV. The undulations in the plot are caused by interference from multiple reflections within the GaN layer. In order to determine $E_2$, the line fitted to the region I is extrapolated to region II. The difference between the measured response and the extrapolated values represents response due to excitations from the InN valence band. The square root of the difference is plotted as a function of photon energy in Fig. 1(c). This is found to be linear and gives the threshold $E_2 = 2.527$ eV.

The interpretation of the measured threshold energies is shown schematically by the energy band diagram of the InN/GaN heterointerface in Fig. 2. The band edge of GaN is observed in this experiment from the rise in the photocurrent at ~ 3.38 eV as shown in Fig.1(a). The difference between the GaN band gap $E_g = 3.38$ eV and the threshold energy $E_2$ directly gives the valence band offset $\Delta E_v = 0.853$ eV. From the threshold energy $E_1$, we get $E_g^{InN} + \xi = E_2 - E_1 = 0.903$ eV, and the conduction band offset $\Delta E_c = E_1 + \xi$ where $\xi$ is the filling of the conduction band with electrons. Using the electron density in InN from Hall measurements, $\xi$ is estimated to be about 0.2 eV[13]. This $\xi$ value matches well with the difference of energy between the room temperature photoluminescence (PL) peak (0.82eV) and the absorption edge (1.02eV) measured for this sample (Fig. 3). This leads to estimated values of $\Delta E_c = 1.824$ eV.

The value of the valence band offset obtained by us is larger than the theoretical values of 0.48 eV obtained by Wei and Zunger[5] and also from the experimental value of 0.5 ± 0.1 eV obtained by Shih et al.[8] from XPS measurements. Considering the high quality of our samples, we feel confident that the heterojunction is relatively free from disorder-related tail states. This is also supported by the PL and absorption measurements. The error involved in the determination of the threshold energy values is estimated to be ~ 0.05 eV. This leads us to conclude that the valence band offset ~ 0.85 eV is a reliable value. The difference between the value determined by Shih et al.[8] may partly arise since very thin layer is used in XPS measurements and the strain in the two samples will be different.

In summary, from the threshold energies in the internal photoemission spectrum of n+ InN on GaN epilayers, we estimate the InN/GaN band offsets to be $\Delta E_v$ = 0.85 eV and $\Delta E_c$ = 1.82 eV.

The authors thank Prof. K. L. Narasimhan for discussions related to internal photoemission. ZHM acknowledges TIFR for experimental facilities and financial support for this work.

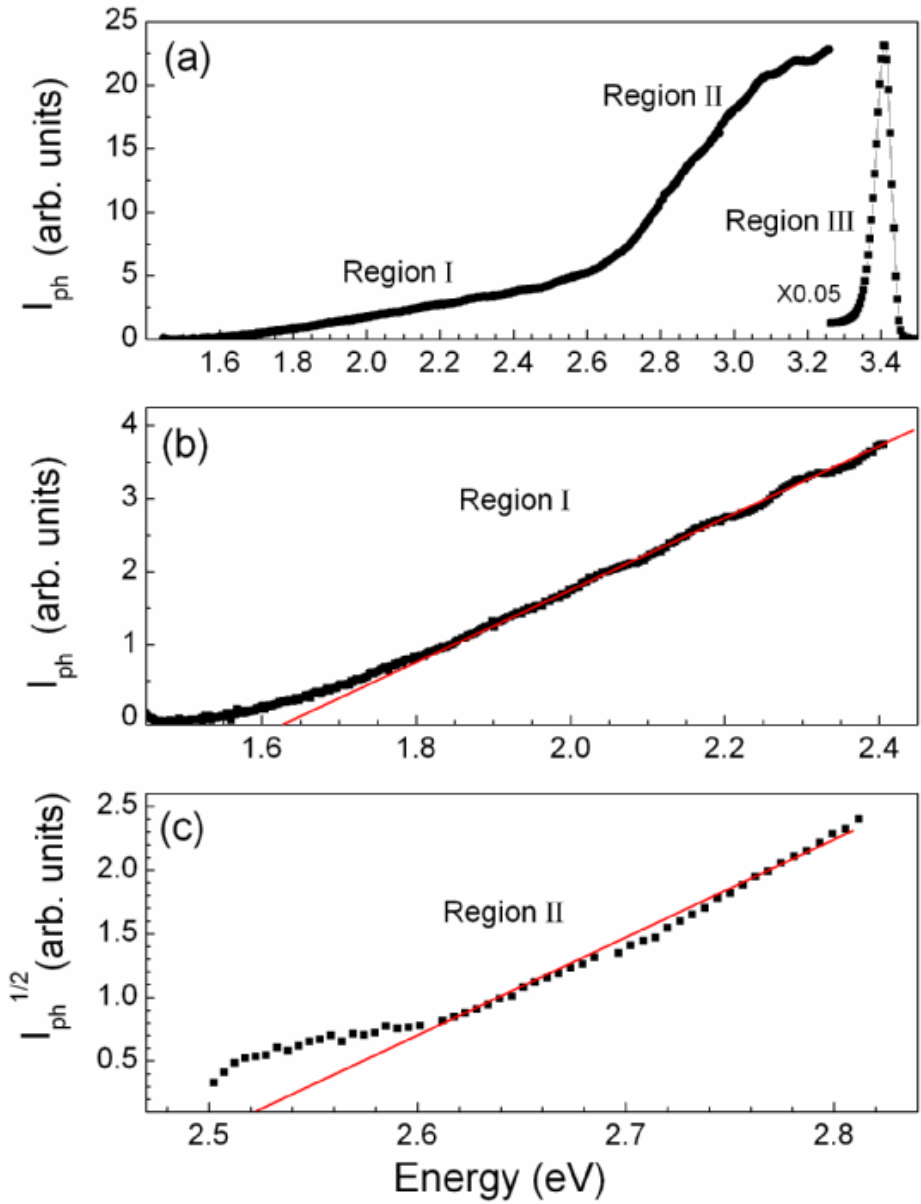

FIG. 1. (a) Normalized photocurrent $I_{ph}$ versus photon energy in the energy range 1.5eV to 3.5eV showing three regions – a linear region (I), nonlinear region (II) and a peak in the energy range of GaN band gap (III). For obtaining the photocurrent spectrum in region (III), the 430 nm long pass filter was removed and spectral band pass reduced to 3.2 nm. (b) Linear fit to the photocurrent in region (I) (c) Square root of the excess photocurrent above the linear response as a function of photon energy in the nonlinear region (II).

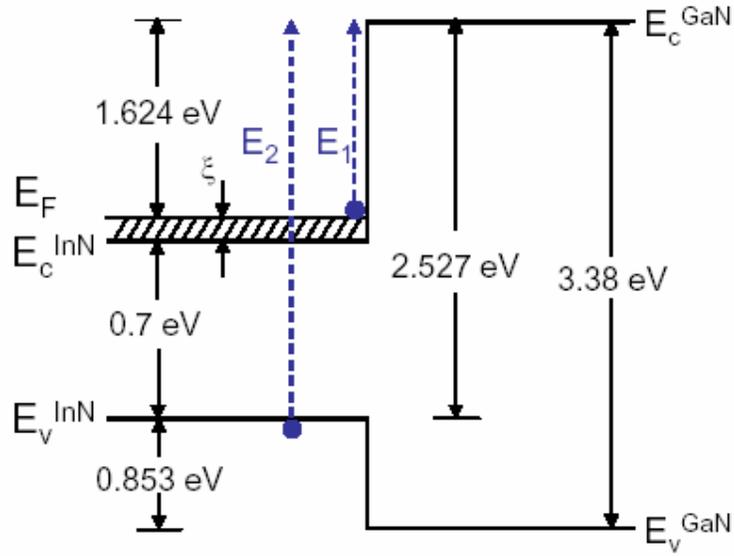

FIG. 2. Schematic model for the band offsets at the InN-GaN heterointerface. Transitions indicated by $E_1$ and $E_2$ correspond to the two threshold energies seen in the photocurrent spectrum. Band bending at the interface has been omitted for simplicity.

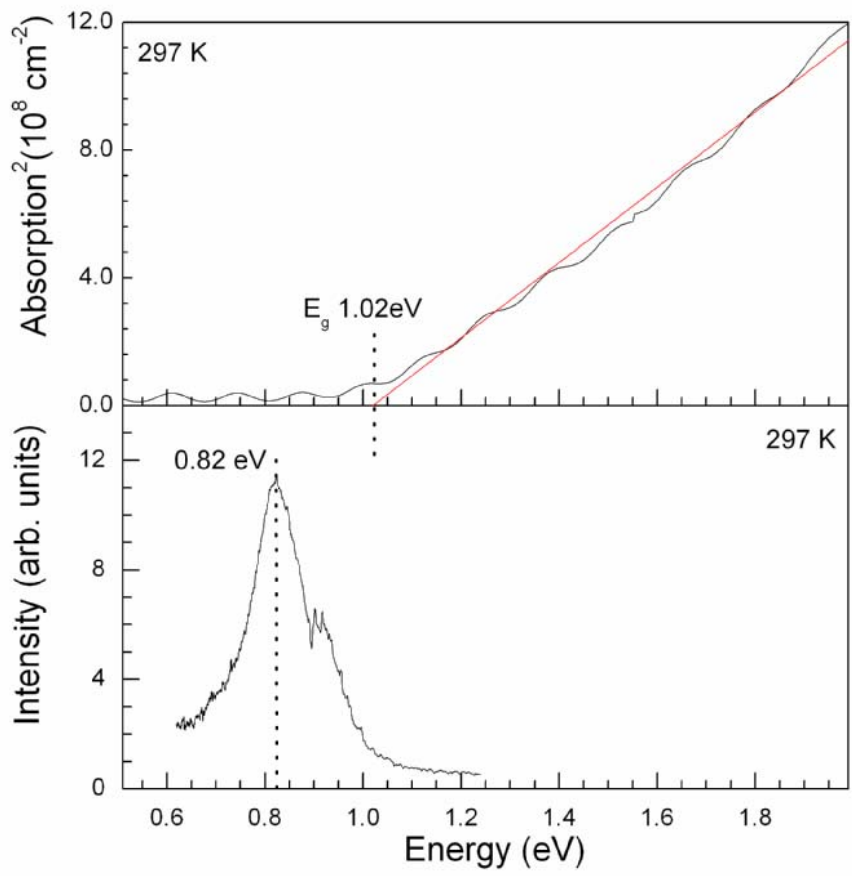

FIG 3. (a) Absorption (plotted as $\alpha^2$) and (b) photoluminescence spectra measured at room temperature on the same sample used for the internal photoemission study.